\documentstyle[12pt,epsf]{article}
\begin{document}

\renewcommand{\thepage}{\arabic{page}}
\setcounter{page}{1}
\newcommand{\mysection}[1]{\setcounter{equation}{0}\section{#1}}
\renewcommand{\theequation}{\thesection.\arabic{equation}}
\newcommand{\nc}{\newcommand}
\nc{\beq}{\begin{equation}} \nc{\eeq}{\end{equation}}
\nc{\beqa}{\begin{eqnarray}} \nc{\eeqa}{\end{eqnarray}}
\nc{\lsim}{\begin{array}{c}\,\sim\vspace{-21pt}\\< \end{array}}
\nc{\gsim}{\begin{array}{c}\sim\vspace{-21pt}\\> \end{array}}

\begin{titlepage}

{\hbox to\hsize{hep-ph/9703246 \hfill EFI-96-50}}
{\hbox to\hsize{December1996 \hfill Fermilab-Pub-97/054-T}}
{\hbox to\hsize{ \hfill to appear in Phys. Lett. B}}
\bigskip

\begin{center}

\vspace{1cm}

\bigskip

\bigskip

\bigskip

{\Large \bf  Some Remarks On Gauge-Mediated Supersymmetry Breaking}

\bigskip

\bigskip

{\bf Erich Poppitz}$^{\bf a}$ and {\bf Sandip P. Trivedi}$^{\bf b}$ \\

\bigskip

$^{\bf a}${\small \it Enrico Fermi Institute\\
 University of Chicago\\
 5640 S. Ellis Avenue\\
 Chicago, IL 60637, USA\\}
 
 \bigskip

$^{\bf b}${ \small \it Fermi National Accelerator Laboratory\\
  P.O.Box 500\\
 Batavia, IL 60510, USA\\}

\bigskip

 {\tt epoppitz@yukawa.uchicago.edu}\\

 {\tt trivedi@fnal.gov}\\ 

\smallskip

\bigskip

\vspace{.8cm}

{\bf Abstract}
\end{center}

We investigate 
the communication of supersymmetry breaking to the Standard 
Model in theories of gauge mediated supersymmetry breaking 
with  general weakly coupled messenger sectors. 
We calculate the one loop gaugino and 
 two loop soft scalar masses for nonvanishing 
${\rm Str} M^2$ of the messenger sector. The 
 soft scalar masses are sensitive to physics at scales higher 
than the messenger scale, in contrast to models with 
vanishing messenger supertrace.
We discuss the implications of this 
ultraviolet sensitivity in theories with 
renormalizable and  nonrenormalizable 
supersymmetry breaking sectors.  We  note that the  standard relation,
in minimal gauge mediation,  between
soft  scalar and gaugino masses   is altered 
 in models with nonvanishing  messenger supertrace.

\end{titlepage}

\renewcommand{\thepage}{\arabic{page}}

\setcounter{page}{1}

\baselineskip=18pt

\mysection{Introduction.}

 In the past couple of years, models where supersymmetry breaking 
 is communicated to the Standard Model by gauge interactions have 
 received increasing attention \cite{oldpapers}-\cite{newpapers}. 
 This development has been stimulated in part by the new  
 advances in understanding the dynamics of $N=1$ supersymmetric 
 gauge theories \cite{seibergexact} and, recently, by the 
 observation of the $e^+ e^- \gamma \gamma$ event at
 Fermilab \cite{CDFevent}.
Gauge mediated supersymmetry breaking offers a  predictive and
 testable  alternative
 to the supergravity models and naturally suppresses flavor changing
neutral currents.
 
 To communicate supersymmetry breaking via gauge interactions one 
 postulates the  existence of heavy vectorlike multiplets of the 
Standard 
 Model gauge group. These heavy ``messenger" fields acquire 
soft supersymmetry
 breaking mass splittings due to their interactions with the 
supersymmetry
 breaking sector. 
 Supersymmetry breaking is then transmitted to the Standard 
Model gauginos,
 which acquire mass at one loop level, and the squarks, 
sleptons and higgses,
 which acquire mass (squares) at the two loop level.

 The interaction of the messenger fields (denoted by $Q_1$ 
and $Q_2$,
 transforming in conjugate representations of the Standard 
Model gauge groups)
 with the supersymmetry breaking sector can be written in 
terms of  a
 ``spurion" field[s] $S$---which is a dynamical field of the 
supersymmetry breaking 
sector---that 
 acquires a supersymmetry breaking vacuum expectation value:
 \beq
 \label{svev}
 \langle S \rangle = s + \theta^2~ F_s ~.
 \eeq 
 The most general interaction\footnote{For simplicity, in 
 eq.~(\ref{interaction}) we impose a (discrete) $R$ symmetry to forbid 
terms like 
 $ S^\dagger Q_1 \cdot Q_2$, etc., in the K\"ahler potential. 
This does not 
 affect the generality of the resulting mass matrix.}
 between the spurion $S$ and the messengers
 $Q_i$, quadratic in the messenger fields,   can be written as 
\cite{girardellogrisaru}:
 \beq
 \label{interaction}
 \int d^4 \theta ~{S^\dagger S \over M^2}~ 
f_i ~Q_i^\dagger \cdot Q_i + 
 \left( ~\int d^2 \theta~
 S~ Q_1 \cdot Q_2 + {\rm h.c.} ~\right) ~.
 \eeq
 The scale $M$ is a scale characteristic of the 
supersymmetry breaking sector.
Inserting the expectation value (\ref{svev}) of 
the spurion generates soft scalar masses
for the messengers and gives rise to the 
 following general scalar mass matrix: 
 \beq
 \label{generalmassmatrix}
 (Q_1^\dagger ~ Q_2)~ \left( \begin{array}{cc} a^2&c^2       \\
          c^{2 *}& b^2
 \end{array} \right)~ \left( \begin{array}{c} Q_1     \\
           Q_2^\dagger
 \end{array} \right)~,
 \eeq
 while the fermion Dirac mass is equal to $s$. 
 The scalar 
 mass matrix in eq.~(\ref{generalmassmatrix}) is the most
 general one can write for a single messenger multiplet. 
 The elements $a, b$ of the mass matrix are real, while $c$ is 
complex; these four real parameters can be related to the   
 parameters in eqs.~(\ref{svev}, \ref{interaction}).

We note that if the coefficients $f_i$ in 
eq.~(\ref{interaction}) are 
not small and if the scale $M$ is not much 
larger than the vacuum expectation value $\langle S \rangle$ 
of the spurion, 
the nonholomorphic scalar masses---the elements $a^2$ 
and $b^2$ in (\ref{generalmassmatrix})---receive
soft supersymmetry breaking  contributions from the D-terms, 
in addition to the (supersymmetric) contribution from
 the F-term. The supertrace of the messenger mass matrix, 
 ${\rm Str} M_{mess}^2 \equiv 2 a^2 + 2  b^2 - 4 s^2$, is then nonvanishing.

 In the models of Dine, Nelson, and Shirman (hereafter referred to as the ``minimal gauge
mediation", MGM models, see ref.~\cite{dnns}), the coefficients $f_i$ in 
 eq.~(\ref{interaction}) are generated by loop effects\footnote{For example, the same
two-loop graphs that generate soft masses for the ordinary squarks and
sleptons also generate corrections to the messengers' soft 
masses---the D-terms 
in eq.~(\ref{interaction})} 
(while $M \simeq s $), 
 and are therefore suppressed---the supertrace of the 
messengers' squared mass
 matrix  vanishes to a good accuracy.
 The models of ref.~\cite{dnns}, however,  require a rather complicated 
structure 
in order
 to give both a supersymmetry preserving  and a supersymmetry
 breaking expectation
 value of the singlet field $S$. Moreover, since the messengers are
 not part of the supersymmetry breaking sector, the 
minimum with the 
required
 supersymmetry breaking expectation value of $S$ is only local.
 Additional complications
 are needed in order to avoid this problem \cite{lisaandlbl}.

It appears natural, therefore, to look for models
where the messengers are an intrinsic part of the supersymmetry
breaking dynamics \cite{dinenelson93}. 
Recently several models of this type have been constructed \cite{PT},
 \cite{berkeley}, \cite{lisa}. 
The supersymmetry breaking sectors of these models are
based in part 
on the $SU(N) \times SU(N-M)$ models, with $M=1,2$ \cite{PST}. 
At low energies, the gauge dynamics in these models can be 
integrated out.
The infrared 
dynamics of the supersymmetry breaking sector is then described
by a {\it weakly coupled} 
nonlinear supersymmetric sigma model, which contains the messenger 
fields (i.e. the fields $Q_i$ above), as well as  
several gauge singlet
fields that  are essential for  supersymmetry breaking\footnote{In these models  the
Standard Model  soft parameters  receive contributions at   energies higher than 
those in  the sigma model  as
well---we will have more to say about these contributions in Section 3.}. 
These Standard Model gauge 
 singlets play the
role of the spurion $S$ in eq.~(\ref{interaction}). The interaction
  between
the ``spurions" and the messengers can be written as in 
(\ref{interaction}),
with the scale $M$ being identified with the scale of the 
vacuum expectation
value, $s$, of $S$. The coefficients $f_i$ are not loop 
suppressed and  the  
supertrace of the messenger fields' mass squared matrix is 
nonvanishing.

For most of the present investigation the detailed dynamics 
of these models is 
not important; we will only use them as an ``existence proof" of 
models with 
weakly coupled messenger sectors  
with nonzero supertrace.  
It is natural to expect that in any dynamical model where the 
messengers participate in the supersymmetry breaking, 
and the low energy dynamics can be described by a weakly coupled
nonlinear sigma model, ${\rm Str} ~M^2_{\rm mess} \ne 0$. This can be
seen by considering the  tree level supertrace mass squared sum rule \cite{WB} 
for a general nonlinear sigma 
model:
\beq
\label{sumrule}
{\rm Str} ~M^2  ~=~ -~ 2 ~R_{i j^*} K^{i l^*} K^{m j^*} W_m W^*_{l^*},
\eeq
where we use the notations of ref.  \cite{WB}: $W_m$ is the 
gradient of the superpotential, $K^{i j^*}$ is the inverse K\" ahler metric, and 
$R_{i j^*}$ is the Ricci tensor of the K\" ahler manifold. In eq.~(\ref{sumrule})
the trace is taken over all states in the sigma model (including the messenger
singlets), however---and explicit examples confirm this---one does not 
expect a restriction of the supertrace on the space of states charged under 
a global symmetry to vanish for a K\" ahler manifold with nonzero curvature.

In the next section we consider in  detail the effect  of the 
nonvanishing
supertrace on the communication of 
supersymmetry breaking to the Standard Model 
fields. We find  
that the two loop scalar soft masses 
are sensitive to physics at momenta higher than the messenger 
scale, in contrast
with the earlier models. This sensitivity can lead to an 
 enhancement or suppression of the
scalar soft masses compared to the gaugino masses.
 In Section 3, we  consider the
implications of this ultraviolet sensitivity 
on the calculability of the soft scalar masses 
in renormalizable and nonrenormalizable
models of supersymmetry breaking. We point out 
 the importance of possible  threshold  (matching) 
contributions of heavy states in the supersymmetry 
breaking sector that carry
Standard Model quantum numbers. Finally, we point out 
the possibility of obtaining superparticle spectra with 
squarks and leptons lighter than the gauginos in the 
``hybrid" models of supersymmetry breaking.
Many of the results reported in this letter have also been obtained by 
 N. Arkani-Hamed, J.  March-Russell, and
H. Murayama,  and reported in $\cite{berkeley}$---we 
thank these authors for discussions.

\mysection{The two loop soft scalar  masses with 
${\bf {\rm Str} ~M^2_{\rm mess} \ne 0}$.}

In this section we will turn to a detailed  calculation of the  
induced soft masses.
Our starting point is the messenger mass matrix  of  eq. ~(\ref{generalmassmatrix})
which when diagonalized is  of the form:
 \beq
\label{massmatrix}
M_S^2  =            \left( \begin{array}{cc} a^2&c^2       \\
          c^{2 *}& b^2
 \end{array} \right)
       \nonumber \\
 =
U \cdot \left( \begin{array}{cc} m_1^2&0 \\0 &m_2^2 
\end{array} \right) \cdot U^
\dagger~,
\eeq
where the unitary matrix $U$ is
\beq
\label{U}
U = \left( \begin{array}{cc} x&\sqrt{1-x^2}~ e^{i \alpha} \\
     - \sqrt{1-x^2}~ e^{- i \alpha} & x\end{array} \right) ~,
\eeq
with $|x| \le 1$.
The Dirac mass of the messenger fermion will be  denoted by $m_f$.

 We note  once again that
the mass matrix above---unlike the MGM case---allows  for the  
supertrace of the
messengers  to be nonvanishing\footnote{In case the messengers belong to
several  different  representations the  corresponding  parameter is the 
supertrace of the 
mass squared matrix weighted by the Dynkin index of the messenger 
representation. 
In  the subsequent discussion 
we will continue to refer to this  loosely as the supertrace of the mass 
matrix. }. 
This feature will in fact play a crucial  role in the discussion below. It is 
also worth mentioning that even for vanishing supertrace the above ansatz
is more general than the one considered in refs.~\cite{martin}, \cite{giudice}.

The calculation of the radiatively induced gaugino and scalar masses
is most easily performed in components. Superfield techniques 
\cite{grs} in theories with broken supersymmetry are generally 
useful for finding the infinite parts of Feynman diagrams---i.e. for calculating 
anomalous dimensions and beta functions---or for calculating finite parts when 
the  mass splittings in the supermultiplets are small. It is in  these 
cases only that the supersymmetry breaking effects 
can be treated as insertions in the relevant Feynman graphs, 
see e.g.  ref.~\cite{yamada}. 
We are however interested in the more
general case when the  supersymmetry breaking splittings 
are not small compared to the supersymmetric messenger mass. 
Treating supersymmetry breaking effects simply as insertions 
is not appropriate  in this case---one needs to use the exact 
superfield propagators in the supersymmetry breaking  background. 
Expressions for the superfield propagators in  
a general supersymmetry breaking background have 
been given in the literature \cite{fullpropagators}. To the best
of our knowledge, only the chiral superfield propagators 
in the supersymmetry breaking background 
have been obtained; even these are prohibitively 
complicated to allow for (two-) loop calculations.

We first consider the contributions  to  the Standard Model gaugino masses.
These arise at one loop. 
\begin{figure}[t]
\centering
\epsfxsize=3.1in
\hspace*{0in}
\epsffile{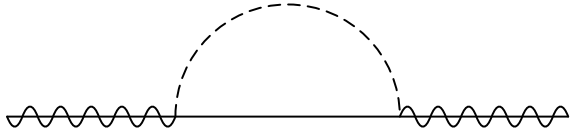}
\caption{One loop messenger 
contribution to Standard Model gaugino masses.}
\end{figure}
The corresponding 
component graph, shown in Fig.~1,  is finite
and the calculation is a straightforward extension of  that in \cite{martin}. 
The result for the  Standard Model gaugino masses is:
\beq
\label{softgaugino}
m_a = e^{-i \alpha}~ x ~\sqrt{1-x^2} ~  {g_a^2 ~m_f \over 4 ~\pi^2} ~S_Q~
{ y_1 ~{\rm log} y_
1 - y_2 ~{\rm log} y_2 -
y_1~ y_2 ~ {\rm log} ( y_1/ y_2 ) \over  (y_1 -1)~(y_2 -1 )}
\eeq
where $y_1 = m_1^2/m_f^2, ~ y_2 = m_2^2/m_f^2$, $g_a$ is the corresponding
Standard Model gauge coupling, and $S_Q$ is
the Dynkin index of the messenger representation (normalized to 1/2 for a
fundamental of $SU(N)$, while for $U(1)_Y$ it is simply $Y^2$, where $Y$ is 
the messenger hypercharge; we do not use GUT normalization of hypercharge).
$m_a$ above refers to the coefficient of the holomorphic gaugino bilinear 
operator $\lambda^{\alpha a} \lambda_\alpha^a$ in the Lagrangian. 

We now turn to a consideration of the scalar masses.  
These arise at two loops\footnote{ 
 We  note that with the more general 
mass matrix (\ref{massmatrix}) 
one loop contributions to the hypercharge D-term are also 
possible. It is easy
to see, however, that these are proportional to $2 x^2 - 1$, and 
therefore
negligible when $x \simeq 1/\sqrt{2}$. Alternatively, if the messengers
fall in complete $SU(5)$ multiplets, these contributions cancel for any
value of $x$. For a generic $x \ne 1/\sqrt{2}$, 
however, one has to also worry about the
two-loop contributions to the hypercharge D-term. Finally, the latter contributions
can be controlled by imposing a discrete symmetry---one could have 
two sets of $SU(5)$ $\bar{5} + 5$ messengers   
 and a  symmetry which exchanges the $5(\bar{5})$  from the first set with 
$\bar{5}(5)$ from the second set. In this case the D-term arises 
at three loops.}.
 As  we will see below when the  supertrace does not vanish 
the full contribution is in fact  ultraviolet divergent.  Uncovering this divergence
though is  subtle  and requires  a careful regularization of the theory.
  
We turn to this issue next.   
We   will  use dimensional reduction 
(DRED) \cite{siegel} to regulate the theory in this
paper.   DRED insures that  the  supersymmetry Ward identities are 
preserved---at least to the two loop  order of the 
calculation \cite{jonesniewenhuizen}.
It also guarantees that the leading divergences cancel
when all two loop graphs contributing to the scalar masses are 
summed up (if one uses conventional dimensional regularization instead, 
even for vanishing supertrace one 
finds that the divergences do not cancel---clearly a result of the fact
that continuation to an arbitrary dimension does 
not preserve supersymmetry).  Recall that in DRED  
  one considers the theory
{\it compactified} to $n = 4 - 2  \epsilon$ dimensions, with 
$\epsilon > 0$.  
Now while  the number of spinor components does 
not change under compactification, 
a vector field  decomposes as an $n$-dimensional vector 
and $2 \epsilon$ scalar multiplets in the adjoint of 
the gauge group---the so-called epsilon-scalars. 
 These  $2 \epsilon$ scalar adjoint multiplets  need to be  fully 
incorporated in DRED for consistency 
 (for a clear introduction and a discussion of
the role of  the epsilon-scalars see ref.~\cite{jonesniewenhuizen}).  

Note that in the theory with broken supersymmetry a mass
term for the epsilon scalars is not
forbidden by gauge invariance.  In fact  the epsilon scalars
receive a divergent  contribution to their mass at one loop 
from the  graphs  shown
in  Fig.~2. 
\begin{figure}
\centering
\epsfxsize=4.1in
\hspace*{0in}
\epsffile{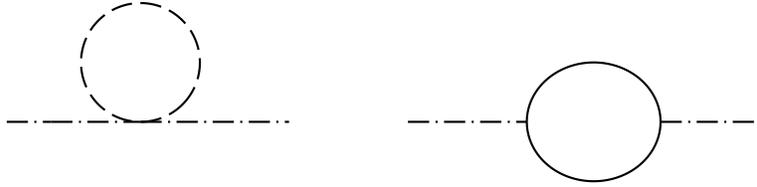}
\caption{Divergent one-loop messenger contributions to the mass of the
epsilon-scalars (dash-dotted lines), proportional to the 
supertrace of the messenger mass
matrix.}
\end{figure}
 This divergent contribution   is proportional to the   
supertrace  of the messenger mass matrix
\cite{jonesmartinvaugnyamada}
and is given by: 
\beq
\label{counter}      
\delta m_{\epsilon}^2 = 
 - S_Q~{g_a^2 \over 16 \pi^2}~
       {{\rm Str} ~M^2_{\rm mess} \over \epsilon}~,
\eeq 
where, as in (\ref{softgaugino}), $S_Q$ is the Dynkin index of the 
messenger representation.
Correspondingly a one-loop counterterm needs to be added to fully 
renormalize the theory---we will choose   the counterterm  to correspond to 
minimal subtraction\footnote{We use the $\overline{DR}^\prime$ 
\cite{jonesmartinvaugnyamada} scheme where no  
``bare" mass for the epsilon scalars is introduced. We thank S. Martin 
for related discussions.}.

Turning  now to the soft  scalar masses of the Standard Model fields,
   one finds that 
this counterterm  contributes to the two loop  soft
 scalar masses via a one loop counterterm graph shown in Fig.~3.
\begin{figure}
\centering
\epsfxsize=2.1in
\hspace*{0in}
\epsffile{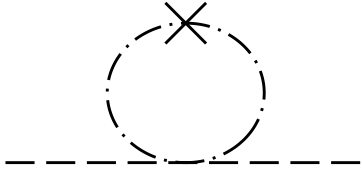}
\caption{One loop epsilon-scalar 
counterterm graph contributing to the two loop 
soft scalar mass.}
\label{countertermgraphm}
\end{figure}
This   resulting  graph is   again logarithmically divergent  
(the factor $\epsilon^{-1}$ in the counterterm corresponding to 
(\ref{counter}) is cancelled
after summing over the 
$2 \epsilon$ adjoint scalar multiplets running in the loop) and gives a 
contribution to the scalar mass of the form: 
\beq
\label{softscalardiv}
m_a^2 = - {g_a^4  \over 128 \pi^4} ~ S_Q ~ C_a ~ 
{\rm Str} ~M^2_{\rm mess}~ {\rm log} {\Lambda_{UV}^2 \over m_{IR}^2} ,
\eeq
where $C_a$ is the quadratic Casimir 
($(N^2 - 1)/2 N$ for an $SU(N)$ fundamental; for $U(1)_Y$
$C_a$ is $Y^2$, with 
$Y$ being the  hypercharge of the Standard Model field involved).
$\Lambda_{UV}$ and $m_{IR}$ refer to the ultraviolet and infrared cutoff
respectively. 

We  now turn to  considering the other graphs at two loops. 
These are identical to the graphs considered in \cite{martin} and 
are shown in Fig.~4.  
\begin{figure}
\centering
\epsfxsize=5.8in
\hspace*{0in}
\epsffile{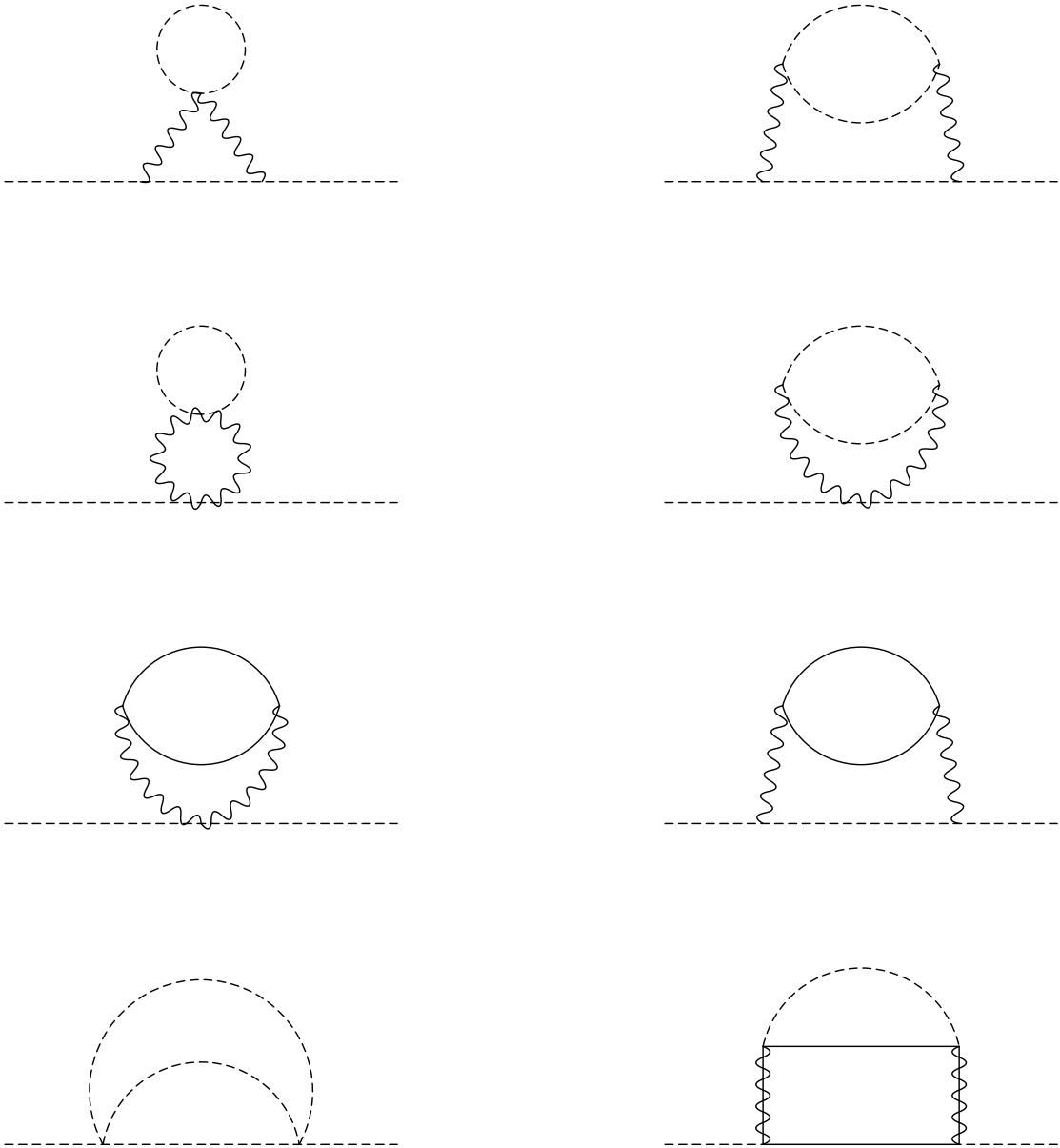}
\caption{Two-loop messenger 
contributions to Standard Model soft scalar masses (the wavy lines denote
 both gauge bosons and epsilon-scalar
propagators). The infrared divergence, present
when Str $M^2_{mess} \ne 0$, is cancelled by the counterterm graph of 
Fig.~3.}
\end{figure}
The contributions of these graphs can be calculated in a manner analogous
to that in \cite{martin}. One finds that these graphs do not give contributions
that are ultraviolet divergent.  They do however give infra-red divergent
contributions and these are cancelled by the infrared divergence in 
eq.~(\ref{softscalardiv}).
In fact it can be argued on general grounds of gauge invariance that  no
infra-red divergences can arise in the soft masses.  
The resulting cancellation between the different contributions is 
therefore a useful check on the calculation.
Putting the contributions from the  graphs in Fig.~3 and Fig.~4 together 
gives finally for the soft  scalar masses\footnote{The
divergent contribution in eq.~(\ref{softscalar}) has previously 
been obtained in refs.~\cite{yamada}, \cite{jonesmartinvaugnyamada}.}:
\beq
\label{softscalar}
m_a^2 = {g_a^4  \over 128 \pi^4} ~ m_f^2 ~ C_a ~ S_Q ~
F( y_1, y_2,  \Lambda_{UV}^2/m_f^2) ,
\eeq
where $C_a$  and   $S_Q$  again refer to the Casimir  and Dynkin indices
respectively as in eq.(\ref{softscalardiv}).  The function $F$ is given by:
\beqa
\label{f}
F( y_1, y_2,  \Lambda_{UV}^2/m_f^2) =&-& (2~ y_1 +  2~ y_2 - 4)~ 
 {\rm log}  {\Lambda_{UV}^2\over m_f^2} \nonumber \\
&+& 2~ (2~ y_1 +  2~ y_2 - 4)+
 4~ x^2 ~(1 - x^2)~(y_1 + y_2) ~{\rm log} 
y_1~{\rm log} y_2  \nonumber  \\
&+& G( y_1, y_2) + G( y_2, y_1) ~,
\eeqa
where
\beqa
\label{g}
G(y_1, y_2)  &=& 2 ~y_1 ~{\rm log} y_1 + 
(1 + y_1) ~{\rm log}^2 y_1 -
2~ x^2 ~(1 - x^2)~(y_1 + y_2) ~ {\rm log}^2 y_1 \nonumber \\
 &+&  2~(1 - y_1)~ {\rm Li}_2 (1 - {1\over  y_1}) +
2  ~(1+y_1)~{\rm Li}_2 (1 - y_1) \\
&-& 4 y_1 ~x^2 ~(1 - x^2)~ {\rm Li}_2 (1 - 
{y_1 \over y_2}) \nonumber~.
\eeqa
As in the discussion of the gaugino masses,  
$y_1 = m_1^2/m_f^2$ and   $y_2 = m_2^2/m_f^2$.  ${\rm Li}_2 (x)$ above
refers to the dilogarithm function and is defined by ${\rm Li}_2 (x)   \equiv
 - \int_0^1 dz z^{-1} {\rm log} ( 1 - x z)$.
It is easy to see that in the limit of vanishing supertrace 
and $x = -1/\sqrt{2}$ eqs.~(\ref{softgaugino}), (\ref{softscalar}) 
reproduce the results of refs.~\cite{martin}, \cite{giudice}.  

As  the first term  in  eq.~(\ref{f}) shows,  in general for a non-vanishing 
supertrace 
of the messenger mass matrix,  the  soft scalar masses will depend on the 
ultraviolet cutoff.  It  is worth noting again that this  ultraviolet divergent
contribution  arises from the one loop counterterm, eq.~(\ref{softscalardiv}). 
  Its presence indicates that in general   the soft masses are sensitive to 
physics at scales higher than the scale of the typical  mass of messengers ("the
messenger scale").  This is to be contrasted 
 with the case of vanishing  weighted supertrace,
when the typical momenta contributing to the scalar 
and gaugino masses are of order the messenger scale.

What the relevant ultraviolet cutoff is, will of course depend on the 
particular model. We will have more to say on this in the next section.
Here we simply note that  if  there is a large hierarchy of scales 
in the supersymmetry breaking sector  leading to  the cutoff  being  much
larger than the messenger scale, the term proportional 
to the supertrace
in eq.~(\ref{softscalar}) is the leading contribution  to the soft 
scalar masses\footnote{ At next to leading order    in this case the 
logarithmically enhanced contributions arising at  three loops
could be  comparable  to the non-logarithmically enhanced  contributions in  
eq.~(\ref{softscalar}).  We have not calculated these three loop  
contributions.}. 
Then the general pattern one observes 
from eq.~(\ref{softscalar})  is that the scalar mass squared is
negative if the  messenger supertrace is positive. Alternatively, 
the soft scalar
mass squared is  positive if the supertrace is negative.
A discussion of the phenomenological relevance of this 
observation is also left for
the following sections. 

{}From the point of view of phenomenology, 
the  main fact of  importance   is that
the relation between the scalar and gaugino soft masses, 
characteristic of the minimal models of gauge mediated 
supersymmetry breaking
\cite{dnns}, no longer holds in models with nonvanishing 
messenger supertrace. 

\mysection{Ultraviolet sensitivity and  phenomenological consequences.}

In this section we   address two main issues.  Section 3.1 
 investigates the
effects that cutoff the logarithmic divergence in the soft 
scalar masses, eq.~(\ref{softscalar}), in the framework of both 
renormalizable and nonrenormalizable models of supersymmetry 
breaking.  Section 3.2, addresses some of the 
phenomenological consequences of models of supersymmetry
breaking with nonvanishing messenger supertrace.

\subsection{What cuts off the logarithm?}

\subsubsection{Renormalizable models.}

The  ultraviolet divergence in the   scalar soft masses
indicates that these  masses are sensitive to short distance physics and cannot be 
fully calculated within the low-energy effective theory.   One needs to therefore go 
beyond the effective theory to  the full underlying  theory to estimate them.  
Roughly speaking one expects  that
 if  in the full theory  there are additional 
 fields  that carry Standard Model quantum numbers and can play the role 
of heavy messengers, and if these heavy fields 
 restore the  full supertrace  to zero,  then  the   logarithmic divergence  would
be cut off by the scale of the heavy messengers  (these heavy
messenger fields could then also contribute  to the masses through threshold
effects).  Whether this happens or not depends on the models under consideration
and it is useful,  in the discussion below,  to distinguish between the case when
the underlying theory is a renormalizable theory and when it is a
nonrenormalizable theory. 
Examples of both types of theories 
of dynamical 
supersymmetry
breaking exist in the literature. 

We first consider the case when the underlying theory is renormalizable. In this 
case
 one can conclude that there must be extra heavy messenger fields  in the full
theory and,  moreover,  that the full  weighted supertrace, after including the heavy
messenger fields, must cancel.  This follows from the following argument: If the full
supertrace does not vanish the  Standard Model soft masses  will continue to be
logarithmically divergent in the full theory and  there would have to be a
counterterm to absorb this divergence.    Since the full theory is renormalizable such
a counterterm would have to be renormalizable as 
 well and  would have to respect
all the symmetries of the  Lagrangian.   Furthermore, 
 this  counterterm   would  involve   a product of the Standard Model matter fields
and  fields from the supersymmetry breaking sector (the "spurion " fields). 
However,  it is easy to see that  no such renormalizable
term can exist---since the soft masses are nonholomorphic, they must 
come from a term in
the K\" ahler potential, and so the counterterm  must necessarily have  
dimension greater
than 4 \cite{girardellogrisaru}.  
 On adding the  contribution of the heavy 
messengers  of mass $m_H$,  the   ${\rm log}(\Lambda_{UV}^2/m_f^2)$  term in 
eq.~(\ref{softscalar})  will be replaced  by
 ${\rm log}(m_H^2/m_f^2)$. 

In addition there could be threshold effects coming from these heavy messengers
as well.  These contributions to the soft scalar mass squares are proportional to 
$(\delta m^2_H/m_H)^2$, where 
$\delta m_H$ is the typical splitting of the heavy supermultiplets.
Whenever the ratio of the mass squared 
splitting, $\delta m^2_H$, of the heavy supermultiplets 
to their mass, $m_H$, 
is  of the same  order as the corresponding
ratio for the light messenger supermultiplets, the
finite contributions of the heavy messengers will be  comparable  to 
 those of the
light messengers (the finite contribution of the heavy messengers
can be additionally enhanced by their multiplicity \cite{PT}, \cite{berkeley}). 
Whether or not such  contributions are 
present is a rather 
model dependent question\footnote{These contributions are present  
e.g. in the $SU(N)\times SU(N-M)$ models (renormalizable
for $N = 4, M = 1$ and $N = 5, M = 2$), where the 
mass squared 
splitting of the
light messenger fields is much  smaller than the supersymmetry
 breaking scale, and the
ratio  $(\delta m^2/m)^2$, is the same for light and heavy 
messenger fields.}.
Generally, however, since the mass splitting of the heavy supermultiplets 
are not expected to be greater than the
supersymmetry breaking scale, $\delta m_H \le M_{SUSY}$, one 
expects that  as the mass  $m_H$ increases, 
the finite contribution $(\delta m^2_H/m_H)^2$
of the heavy messengers becomes negligible.

\subsubsection{Nonrenormalizable models.}
 
We now turn to discussing nonrenormalizable 
supersymmetry breaking sectors. 
These typically contain interactions suppressed by 
some scale $M_{UV}$, and are themselves effective field 
theories valid below that scale.  
For the  nonrenormalizable model to be a useful starting point in 
calculating the vacuum expectation values involved
we need that both  
$s$ and $F_s$ are $<< M_{UV}$. Here $s$ and $F_s$  are   the
vevs of the spurion field  eq. (\ref{svev}) and  represent the typical expectation 
values in the supersymmetry breaking sector. 

Eqs.~(\ref{svev}),~(\ref{interaction}), 
and (\ref{generalmassmatrix}) show that one
expects the  leading order contributions  to the supertrace   to  be  $ \sim
(F_s/s)^2$.   In addition  there could be  subleading contributions which 
go like $ (F_s/M_{UV})^2$. We now argue that the leading order 
contribution $\sim (F_s/s)^2$  must  vanish  when we include all the fields in 
the
theory below the scale $M_{UV}$.  In the previous section we used 
renormalizability  to
argue  for the absence  of a possible counterterm and therefore for the vanishing of
the supertrace. This argument is not directly applicable here, since we begin with a
nonrenormalizable model.   
Note though, that while nonrenormalizable
counterterms might be allowed, they must still 
be polynomial in momenta and masses 
(the latter in the  present context are  dynamically generated).  
But a little thought shows that there is no polynomial
counterterm\footnote{Nonpolynomial 
 counterterms  can  be  written---e.g. the D-term
$\Phi^\dagger \Phi~ |{\rm log} S|^2$, where $\Phi$ is a Standard
Model field and $S$ a supersymmetry breaking sector ``spurion" field 
with expectation value (\ref{svev}).}
 that can account for the   log-divergent
contribution to the soft masses
of the form (\ref{softscalar}) when Str $M^2 \sim (F_s/s)^2$. 
 Hence,  we conclude that in
nonrenormalizable models,   the leading 
logarithmic divergence is  cutoff  by  heavy  messenger fields  with a mass 
$m_H$, 
$\Lambda \le m_H < M$ ($\Lambda$ here is the scale above which
 the sigma  model that
describes the light messengers breaks down). 

Besides the leading contribution to the supertrace though there can be, as was
mentioned above, subleading contributions, proportional to $(F_s/M_{UV})^2$.
These  do not have to vanish---the corresponding  polynomial counterterm is of 
the form $ \Phi^\dagger \Phi  S^\dagger S/M_{UV}^2$.
The importance of these subleading contributions to the soft scalar mass
in any model will depend on the ratio of the scales $s/M_{UV}$, the 
magnitude of the 
logarithmic enhancement, and the coefficients of the counterterms mentioned 
above. 

\subsection{Phenomenological consequences.}

As noted at the end of Section 2, the logarithmically
enhanced term in eq.~(\ref{softscalar}) changes the relation
between scalar and gaugino soft mass parameters, $m_{gaugino} \sim
m_{scalar}$, typical of models with vanishing supertrace. 
The logarithmic contribution
to the scalar soft masses is expected to dominate over 
the finite contribution---the finite threshold corrections 
due to both heavy and light messenger fields---in case 
there is a sufficiently large hierarchy of 
scales in the supersymmetry breaking sector.  
As we now discuss, this in fact puts a significant constraint on 
the class of viable models.  For, 
as eq.(\ref{softscalar}) shows,   when the supertrace  for the 
light messengers  is positive the logarithmically enhanced  term provides 
a negative contribution to the scalar mass squares.  Therefore when this term 
dominates the scalars  are driven  to acquire vaccum expectation values 
and,  in particular,  $SU(3)_c \times U(1)$
is broken---a clearly unacceptable  outcome\footnote{Similar observations were 
made in  \cite{berkeley}.}.

One obvious  way to try and avoid this  possibility  would be to construct
 models 
where the supertrace of the light messengers is negative.  
The models  with supersymmetry
breaking-cum-messenger sectors  that have been  studied in detail  so far, 
in particular the models of  \cite{PT}, \cite{berkeley}, have
all yielded  a   positive supertrace of the light
messengers\footnote{We note that  L. Randall has recently constructed
some models that have a negative supertrace of the messengers \cite{lisa}.}.  
This poses a serious problem for these models.  In 
  \cite{PT}, for example,  obtaining  positive scalar mass squares  requires,
for $11 \le N \le 27$, 
the scale $m_H$ to be $2.4 -2.5$ times the light messenger 
scale. This  is  clearly an unsatisfactory situation in 
which case the weak coupling analysis
of the ground state is not even valid.
We are however not aware of any general argument that requires  the positivity
of the light messenger supertrace\footnote{It is easy to construct
nonlinear sigma models that incorporate supersymmetry breaking and 
light messengers, in which the sign and magnitude of the supertrace is
a free parameter (for example, in the $SU(N)\times SU(N-2)$ models
this can be achieved by adding additional terms, allowed by all
symmetries, to the K\" ahler potential of ref.~\cite{PT}). However, we are
not aware of any dynamical models to which these sigma models are
a consistent low-energy approximation. }.  This possibility 
might even be realized
in a  more exhaustive study of other vacua of known models. 
Finally, it is worth mentioning that even in models with the required sign for the
supertrace, the logarithmically enhanced term might still pose a problem
by driving the scalars much heavier  and thus resulting in light gluinos.

In fact the logarthmic term in eq.~(\ref{softscalar}) can be quite  significant even when 
the ratio of the heavy to light messenger scales is not very large. 
To illustrate this consider an example consisting of two sets of messengers
(both, say, in the fundamental representation of the relevant group).
Further let the light messenger fields have a non-vanishing supertrace 
which is cancelled by the supertrace of the heavy fields thereby 
setting the full supertrace to zero. On choosing the ratio of the 
heavy to the light messenger fermion masses to be $\sim 3$ and choosing
the light supertrace to be positive $\simeq $ the light fermion mass in 
magnitude, one finds that the Standard Model scalar squared 
masses are generally 
negative. Further, if the sign of the light supertrace is reversed to be 
negative, the Standard Model masses now become generally positive. 
The logarithmic term can 
thus have a significant effect even for a small separation of scales
 between the heavy and light messengers and obtaining positive 
soft scalar masses in its presence is a significant constraint.

We  conclude this  section by  briefly  commenting on the "hybrid" models of 
\cite{PT}.  In these models  the soft scalar masses get comparable contributions
from  both gauge and gravitational effects. The negative  contributions to  scalar 
masses  (arising from a positive light supertrace)  from gauge mediation are then not
necessarily a problem since they could be compensated by positive supergravity
contributions.   In fact  they could lead to squarks and sleptons being lighter than
gluinos---a novel  and quite  distinct spectroscopy\footnote{We thank G.
Anderson  for discussions in this regard.}.  For example,  one can check that in the
$SU(17)\times SU(15)$  models the (positive) supergravity contribution to the soft
masses (due to the  term that cancels the cosmological constant and to higher
dimensional terms in the K\" ahler potential) is comparable to the
(negative) log-enhanced 
contribution of the light messengers' supertrace. A fortuitous cancellation
between these two contributions  could then lead to squarks which are generically
lighter than the gauginos. It would be interesting to study the renormalization
group effects in these  models in some detail---superparticle spectra with squarks
much lighter than gauginos can not arise in (minimal) supergravity models.

We acknowledge useful discussions with Bill Bardeen, Hsin-Chia Cheng,  
 Keith Ellis,  
Steve Martin, John March-Russell, Yael Shadmi, and especially Greg Anderson.
E.P. acknowledges support by DOE contract DF-FGP2-90ER40560.
 S.T. acknowledges
the support of DOE contract DE-AC02-76CH0300.

\nc{\ib}[3]{ {\em ibid. }{\bf #1} (19#2) #3}
\nc{\np}[3]{ {\em Nucl.\ Phys. }{\bf #1} (19#2) #3}
\nc{\pl}[3]{ {\em Phys.\ Lett. }{\bf #1} (19#2) #3}
\nc{\pr}[3]{ {\em Phys.\ Rev. }{\bf #1} (19#2) #3}
\nc{\prep}[3]{ {\em Phys.\ Rep. }{\bf #1} (19#2) #3}
\nc{\prl}[3]{ {\em Phys.\ Rev.\ Lett. }{\bf #1} (19#2) #3}

\end{document}